\documentclass[twocolumn,prb,superscriptaddress,showpacs,preprintnumbers,amsmath,amssymb]{revtex4}
\usepackage{graphicx}% Include figure files
\usepackage{dcolumn}% Align table columns on decimal point
\usepackage{bm}% bold math

\begin{document}

\preprint{APS/123-QED}

\title{Optical pumping NMR in the compensated semiconductor InP:Fe}

\author{Atsushi Goto}
 \email{GOTO.Atsushi@nims.go.jp}
 \affiliation{National Institute for Materials Science, Sakura, Tsukuba, 305-0003, Japan}

\author{Kenjiro Hashi}%
 \affiliation{National Institute for Materials Science, Sakura, Tsukuba, 305-0003, Japan}

\author{Tadashi Shimizu}
 \affiliation{National Institute for Materials Science, Sakura, Tsukuba, 305-0003, Japan}

\author{Ryo Miyabe}
 \thanks{Present address: Furukawa Electric Co. Ltd.}
 \affiliation{National Institute for Materials Science, Sakura, Tsukuba, 305-0003, Japan}
 \affiliation{Department of Applied Physics, Tokyo University of Science, Shinjuku, Tokyo 162-8601, Japan}

\author{Xiaogang Wen}
 \thanks{Present address: National Institute of Advanced Industrial Science and Technology.}
 \affiliation{National Institute for Materials Science, Sakura, Tsukuba, 305-0003, Japan}

\author{\\Shinobu Ohki}
 \affiliation{CREST, Japan Science and Technology Agency, Honcho, Kawaguchi, Saitama 332-0012, Japan}

\author{Susumu Machida}
 \affiliation{Quantum Entanglement Project, JST, 3-9-11 Midori-cho, Musashino, Tokyo 108-0012, Japan}

\author{Takahiro Iijima}
 \affiliation{National Institute for Materials Science, Sakura, Tsukuba, 305-0003, Japan}

\author{Giyuu Kido}
 \affiliation{National Institute for Materials Science, Sakura, Tsukuba, 305-0003, Japan}
 \affiliation{Department of Applied Physics, Tokyo University of Science, Shinjuku, Tokyo 162-8601, Japan}

\date{\today}% It is always \today, today, but any date may be explicitly specified

\begin{abstract}
The optical pumping NMR effect in the compensated semiconductor InP:Fe has been investigated
in terms of the dependences of photon energy ($E_p$), helicity ($\sigma^{\pm}$), 
and exposure time ($\tau_{L}$) of infrared lights.
The $^{31}$P and $^{115}$In signal enhancements show large $\sigma^{\pm}$ asymmetries 
and anomalous oscillations as a function of $E_p$.
We find that
(i) the oscillation period as a function of $E_p$ is similar for $^{31}$P and $^{115}$In
and almost field independent in spite of significant reduction of the enhancement in higher fields. 
(ii) A characteristic time for buildup of the $^{31}$P polarization under the light exposure 
shows strong $E_p$-dependence, but is almost independent of $\sigma^{\pm}$.
(iii) The buildup times for $^{31}$P and $^{115}$In are of the same order ($10^3$ s),
although the spin-lattice relaxation times ($T_1$) 
are different by more than three orders of magnitude between them.
The results are discussed in terms of 
(1) discrete energy spectra due to donor-acceptor pairs (DAPs) in compensated semiconductors, 
and (2) interplay between $^{31}$P and dipolar ordered indium nuclei, 
which are optically induced. 
\end{abstract} 

% PACS, the Physics and Astronomy Classification Scheme.

\maketitle

\section{\label{sec:introduction}Introduction}
A solid-state NMR quantum computer has attracted much attention 
because of its great potential as a scalable quantum computer,
\cite{divincenzo95,yamaguchi99,goto02a,ladd02,goto02b,goto03a}
but it holds the problem of extremely low efficiency 
in initialization and readout processes 
due to low nuclear spin polarization at thermal equilibrium. 
One of the possible resorts to resolve it is an optical pumping NMR in semiconductors,
\cite{shimizu00,ladd02,goto03a}
which provides an aligned nuclear spin system inside a semiconductor 
through transfer of angular momenta from photons to nuclei via electrons.
\cite{lampel68,meier84}
The aligned nuclei can be used as an initial state of the pseudo-pure technique for initialization
\cite{gershenfeld97}
as well as for a signal enhancement for readout.
Moreover, such a semiconductor can be used as a reservoir of the spin polarization,
which is transferred to nuclear spins in another material that serves as a quantum computer.
\cite{tycko98}
The scheme is specifically called an
\textit{optical pumping qubit initializer (OPQI)}.\cite{goto03a,goto03c}
It allows us to separate initialization process from computation,
so that the latter can be optimized independently of the former.

The OPQI requires a polarizer (reservoir) with the following characteristics;
(1) high optical pumping efficiency, 
(2) large nuclear moment and 
(3) high spin transfer efficiency from the polarizer to a polarized material.\cite{goto03c,tycko98}
Among many semiconductors, indium phosphide (InP) is expected to possess preferable characteristics 
at least for (2) and (3),
i.e., $^{31}$P has 100\% abundance and a high nuclear gyromagnetic ratio $\gamma_n/2\pi$ = 17.235 MHz/T, 
which enables us to retain high nuclear spin moments.
Moreover, since the nuclear spin of $^{31}$P ($I$) is 1/2,
no quadrupolar broadening of the $^{31}$P spectrum exists at an interface,
which reduces a chance of degrading the polarization transfer efficiency.
 
It is known, however, that InP has problems in (1);
it shows rather complicated optical pumping effects,
which prevents us from optimizing the enhancement of $^{31}$P polarization.
It shows strong dopant dependence, 
and clear enhancement of a $^{31}$P NMR signal has been reported only in undoped\cite{tycko98,patel99} 
and Fe-doped samples.\cite{michal99,hashi03,goto03b}
Moreover, a Fe-doped sample, which shows the most significant enhancement,
exhibits rather peculiar and unique optical pumping behaviors
such as,
(i) an oscillatory behavior of the $^{31}$P enhancement against a photon-energy ($E_p$) of an incident light,
(ii) an asymmetric enhancement in terms of a photon-helicity 
(right- and left-circular polarizations, $\sigma^{\pm}$),\cite{michal99}
and (iii) single resonance polarization transfer from $^{115}$In to $^{31}$P with only a rf-field for $^{31}$P.
\cite{michal98}
In order to optimize the $^{31}$P enhancement in InP, 
it is inevitable to understand the mechanism of these behaviors 
and gain information about guiding principle for an effective optical pumping.

In this paper, we report on optical pumping effects in InP:Fe
in terms of $E_p$, $\sigma^{\pm}$ and light exposure time ($\tau_{\rm L}$) dependences
under the various experimental conditions of magnetic field and temperature.
The results are discussed in terms of the unique characteristics of this system, 
with which the conditions for the effective optical pumping in InP are discussed.

\section{\label{theory}Optical pumping NMR in III-V semiconductors}
\begin{figure}[tb]
\includegraphics[scale=0.5]{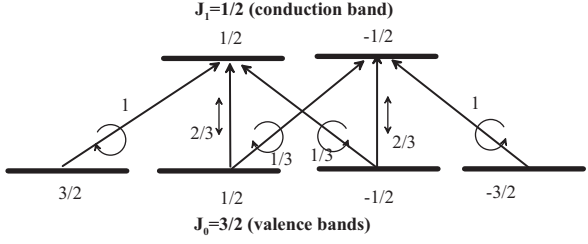}
\caption{\label{selection}
Diagram of the allowed inter-band transitions at the $\Gamma$ point of III-V semiconductors. 
The numbers below (above) the initial (final) states 
indicate the z-components of the total angular momentum, $J_z$.
The helicities/polarities and the relative transition probabilities are shown 
by circular/linear arrows and the numbers next to them, respectively.}
\end{figure}
Here, we summarize an optical pumping NMR in III-V semiconductors.
Primarily, the optical pumping effect is caused by a transfer of angular momenta 
from incident photons to nuclei via electrons.
Photons with $\sigma^{\pm}$ have the angular momenta $l=\pm1$,
which pump up the electrons from the heavy- and light-hole bands to the conduction band 
by adding (subtracting) an angular momentum of $\pm1$ to the electron spins, 
as shown in Fig. \ref{selection}.\cite{meier84}
Due to the difference in the transition probabilities,
the excess electrons with either spin-up or down states are created by $\sigma^{\pm}$,
depending on the sign of the electron g-factor.
The z-component of the photoexcited electron spin thus created 
is relaxed with an electron spin relaxation time $\tau_s$.
In the steady state, a net spin $S_z$ is given by,
\begin{equation}
S_z=\frac{S_i}{1+\tau/\tau_s},
\label{S}
\end{equation}
where $S_i$ corresponds to $S_z$ at the instant of the photon-absorption,
and $\tau$ is the lifetime of the photoexcited electrons.

The spin polarization $s \equiv \langle S_z \rangle/S(S+1)$ is transferred to nuclei.
($\langle \: \rangle$ 
stands for an spatial average over a region large enough compared to the lattice spacing $a$.)
The nuclear spin polarization $p\equiv\langle I_z \rangle/I(I+1)$ is given by,
\begin{equation}
 p=\frac{s-s_0}{1-4ss_0},
\label{Iav}
\end{equation}
where $s_0=S_0/S(S+1)$ with 
\begin{equation}
S_0=-\frac{1}{2} \tanh(\frac{\mu_BgB}{2k_{\rm B}T})
\end{equation}
being the electron spin at thermal equilibrium at $T$.
In the optical pumping experiments, 
\begin{equation}
 p=s
\label{steady1}
\end{equation}
because $s_0 \ll s$,
i.e., the average nuclear polarization is proportional to that of the electron spins.
Note that $p=-s_0$ in the Overhauser effect because $S_z=0$.

The dynamics of the polarization is given through hyperfine couplings 
with the pumped electrons trapped in trapping sites.
The light exposure time $\tau_{\rm L}$ dependence of $p$ is given by,\cite{kuhns97,patel99}
\begin{equation}
 p(\tau_{\rm L}) = p_{\infty}[1-\exp\{-\tau_{\rm L}(1/T_1+1/T_{II})\}],
\end{equation}
where $T_1$ is the spin-lattice relaxation time in the dark, 
and $T_{II}$ is the cross relaxation time between electrons and nuclei.
In the case of a Fermi contact type interaction, $T_{II}$ is given by,\cite{kuhns97,patel99}
\begin{equation}
 \frac{1}{T_{II}}=\frac{S(S+1)}{3}\cdot\frac{2A^2\tau_e}{1+(\omega_I-\omega_S)^2\tau_e^2}.
\label{TII}
\end{equation}
Here, $A$ is a hyperfine coupling constant, 
$\tau_e$ is a correlation time for exchange of electrons between the conduction band and the trapping sites, 
and $\omega_I/2\pi$ and $\omega_S/2\pi$ are the nuclear and electron Larmor frequencies, respectively.
Since $\omega_I/\omega_S \ll 1$ and S=1/2, Eq. (\ref{TII}) is reduced to,
\begin{equation}
 \frac{1}{T_{II}} \approx \frac{A^2\tau_e}{2(1+\omega_S^2\tau_e^2)}.
\label{TII-2}
\end{equation}
A contact type interaction is expected to be dominant in the case of shallow donors,
because polarized electrons directly interact with nuclei of order $10^3$,
and because $\tau_e$ in Eq. (\ref{TII}) is short due to 
the frequent hopping of electrons between the conduction band and the trapping sites.
In the case of deep impurity centers, on the other hand, the carriers are localized so that 
a dipolar type hyperfine coupling rather than the contact type plays a dominant role. 
In this case, non-secular terms in the dipolar hyperfine coupling such as $I^+S_z$ and $I^-S_z$ are dominant.
Since no spin flips of electrons are needed in this process, 
the energy mismatch between $\omega_S$ and $\omega_I$ is of no importance.  

The nuclear polarizations, created either by the Fermi contact or the dipolar hyperfine couplings, 
are transmitted to the region far from the impurity center
via the spin diffusion process with homo-nuclear dipolar interactions.
In the dilute limit, the diffusion process is described by,
\begin{equation}
 \frac{\partial p}{\partial t}=D\Delta p-\eta\frac{s-s_0}{T_{II}}-(p-p_0)\left \{\frac{1}{T_1}+\frac{1}{T_{II}}\right \},
 \label{diffusion}
\end{equation}
where $p_0=\langle I_0 \rangle/I(I+1)$ is a spin polarization at thermal equilibrium, 
which is small compared to $p$ in the optical pumping process.
$D$ is a spin diffusion constant caused by homo-nuclear dipolar couplings,
which is of the order of $Wa^2$ 
with $W$ being the probability of a flip-flop process between nearest homo-nuclear spins.
$\eta$ is a numerical factor, which is 1 for the Fermi contact case and 1/2 for the dipolar coupling case.
If $D$ is neglected and $\eta=1$, the steady state solution of Eq. (\ref{diffusion}) is given by,
\begin{equation}
 p_{\infty}=\frac{s-s_0}{1+T_{II}/T_1},
 \label{steady2}
\end{equation}
If a leakage by $T_1$ can be ignored ($T_1 \rightarrow \infty$), 
Eq. (\ref{steady2}) is reduced to Eq. (\ref{steady1}).

\section{\label{sec:experimentals}Experimental details}
In this section, we describe the experimental setup used in this study.\cite{goto03b}
A Ti:Sapphire tunable laser pumped by a diode-pumped Nd:YVO$_4$ cw green laser
is used to produce a linearly polarized light with the wavelength between 850 and 1050 nm.
The lights are delivered to the tip of an NMR probe in a cryostat installed
in superconducting magnets (6.347 T and 11.748 T) 
by a \textit{polarization-maintaining} (PM) single mode optical fiber (Fujikura), 
which transmits a light with its linear polarization maintained.  
A quarter-wave plate is attached to the tip of the NMR probe, 
which converts the linearly polarized light to the circular one.
The helicity of the circularly polarized light ($\sigma^{\pm}$) can be changed by a half-wave plate 
located on the optical table before the light is introduced into the fiber.

The samples used were commercially available InP wafers doped with Fe (Showa Denko Co. Ltd.).
We also used three other samples (nominally undoped, and doped with S, Zn) as references.
The InP wafers 350 $\mu$m thick
were cut into 5 $\times$ 8 mm rectangular shapes and mounted on the NMR probe
with the surfaces (100) normal to the magnetic field.
The samples were wound by a copper wire, which served as a detection coil.
The circularly polarized lights were illuminated to the samples through the aperture of the coil.
The diameter of the illuminated spot was about 5 mm.
NMR signals of $^{31}$P were detected at temperatures between 4 and 40 K
by a pulsed NMR spectrometer
with the pulse sequence of comb (64 $\times \pi/2$-pulses)-$\tau_{\rm L}$-$\pi/2$ pulse-FID,
where the first 64 comb pulses extinguished a nuclear magnetization, 
and that built up by the infrared lights during the exposure time of $\tau_{\rm L}$
was detected by a free induction decay signal.
The {\it dark} experiment was performed with the same pulse sequence but without laser irradiation,
where $\tau_{\rm LD}$, instead of $\tau_{\rm L}$, corresponds to the long delay time 
in the saturation recovery experiments.

\section{\label{sec:results}Results}
\subsection{Roles of dopants}
Before detailed experiments on InP:Fe,
we examined the optical pumping effects of $^{31}$P in InP with four different dopants, 
which are summarized in Table \ref{table:1}
along with their basic properties provided by the manufacturer.
Among the four dopants, only Fe shows a clear optical pumping effect.
The nominally undoped sample shows a weak effect, 
while little effects are observed in the S and Zn doped samples.
Also shown in Table \ref{table:1} are $T_1$'s of $^{31}$P in the dark, 
which exhibits a correlation with the optical pumping effect,
i.e., more intense effect is observed for the dopants with longer $T_1$ of $^{31}$P.
The qualitative explanations for these results can be given as follows; 
\begin{table}[t]
\caption{\label{table:1}Properties and the optical pumping effects of $^{31}$P of the InP samples used in the present study.} 
\begin{tabular}{@{\hspace{\tabcolsep}\extracolsep{\fill}}ccccc} \hline \hline
dopant \hspace{2mm}& carrier\hspace{2mm} & $\rho$ (cm$^{-3}$)\footnote{\ Carrier density at 300 K.}\hspace{2mm} & $T_1$ (s)\footnote{\ $T_1$ of $^{31}$P at 4.2 K.}\hspace{2mm} & optical pumping\\ \hline
Zn      &  p & $ 5 \times 10^{18} $ & 100 & no\\ 
S       &  n & $ 6 \times 10^{18}$ & 130 & no\\ 
undoped &  n & $ 5 \times 10^{15}$ & 1300 & weak \\ 
Fe      &  S-I\footnote{\ S-I: Semi-insulator} & $ 7 \times 10^{7}$ & $> 10^4$ & intense\\
\hline \hline
\end{tabular}
\end{table}

(1) Zinc-doped sample: $p$ type carriers with relatively high density.
In general, $\tau_s$ of holes in the valence bands is very short
because a strong spin-orbit coupling makes $\tau_s$ coupled tightly to the momentum relaxation time $\tau_p$.
Moreover, exchange interactions between holes and electrons 
cause a short $\tau_s$ of the pumped electrons (Bir-Aronov-Pikus mechanism).\cite{meier84}
Eq. (\ref{S}) predicts that the short $\tau_s$ gives rise to little optical pumping effects.

(2) Sulfur-doped sample: $n$ type carriers with relatively high density.
This sample has a relatively short $T_1$ due to the hyperfine interaction of a Fermi contact type 
between trapped electrons and $^{31}$P.
The spin polarization of $^{31}$P leaks away at the rate of $T_1$.
This results in $T_{II}/T_1 \gg 1$ and $p$ is reduced to zero, as indicated in Eq. (\ref{steady2}).

(3) Nominally undoped sample: $n$ type carriers with relatively low density.
 The $T_1$ of $^{31}$P is relatively long due to the low carrier density, 
and a weak optical pumping effect is observed.

(4) Iron-doped sample: a compensated semiconductor whose carrier density is extremely small.
The low carrier density causes a very long $T_1$ 
and the polarization transferred from the optically pumped electrons are accumulated in the $^{31}$P nuclei.

\subsection{\label{op-InP:Fe}Optical pumping NMR in InP:Fe}
\subsubsection{Temperature and light exposure time dependences of the $^{31}$P signal enhancement}
\begin{figure}[t]
\includegraphics[scale=0.39]{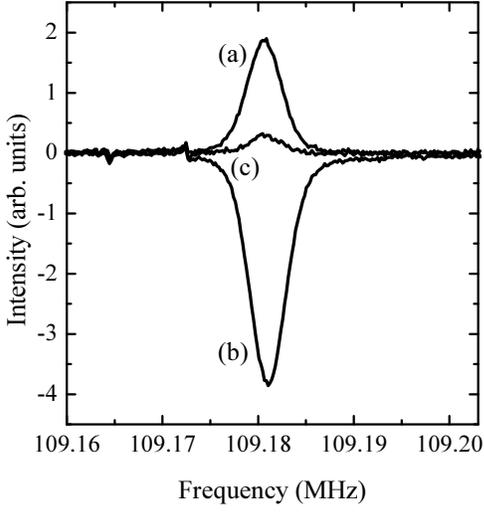}
\caption{\label{spectra}
Optically pumped $^{31}$P NMR spectra 
at $T=4.2$ K and $H_0=6.347$ T with $\Phi=430$ mW/cm$^2$ and $\tau_{\rm L}=300$ s.
(a) $\sigma^-$ and $E_p=1.420$ eV, 
(b) $\sigma^+$ and $E_p=1.416$ eV. 
(c) A spectrum in the dark case with $\tau_{\rm LD}=300$ s.}
\end{figure}

In what follows, we will restrict ourselves to the case of InP:Fe,
where the most intense optical pumping effect is observed.
Figure \ref{spectra} shows the $^{31}$P NMR spectra of InP:Fe under light irradiation
(power density $\Phi$=430 mW/cm$^2$), together with that in the dark case. 
The absolute value of the integrated spectral intensity is enhanced for either helicity, 
but the phase of the signal for $\sigma^+$ is shifted by 180$^o$ (negative) from the dark case.
The negative enhancement for $\sigma^+$ suggests that 
a hyperfine coupling is of dipolar origin.\cite{patel99}

\begin{figure}[t]
\includegraphics[scale=0.32]{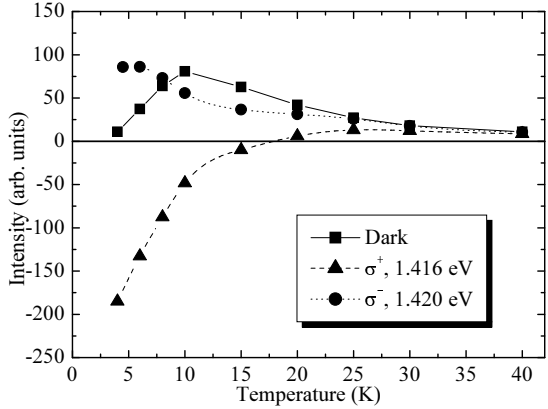}
\caption{\label{temperature}
Temperature dependence of the integrated $^{31}$P signal intensity 
with $\Phi$=430 mW/cm$^{2}$, $\tau_{\rm L}=600$ s at $H_0=6.347$ T. 
$E_p$=1.420 eV and 1.416 eV for $\sigma^-$ and $\sigma^+$, respectively. 
The temperature dependence of the integrated intensity for the dark case with $\tau_{\rm LD}=600$ s is also shown.
}
\end{figure}

%\subsubsection{Temperature dependence}
Figure \ref{temperature} shows the temperature dependence 
of the integrated $^{31}$P signal intensity with $\Phi$= 430 mW/cm$^2$ and $\tau_{\rm L}=600$ s, 
together with that in the dark case with $\tau_{\rm LD}=600$ s. 
The enhancement is more significant at lower temperature
because of a long $\tau_s$ there.
In general, it is expected that $\tau_s$ decreases by a few orders of magnitude from 4.2 K to 300 K,
while the lifetime $\tau$ increases only slightly in the same temperature range.
From our data, the crossover temperature, above which $\tau_s/\tau > 1$ and $S \rightarrow 0$ in Eq. (\ref{S}), 
is estimated to be about 30 K.
A peak around 10 K observed in the dark case is due to $T_1$, 
which increases quite rapidly below 10 K.

%\subsubsection{Buildup time of $^{31}$P polarization}
\begin{figure}[t]
\includegraphics[scale=0.4]{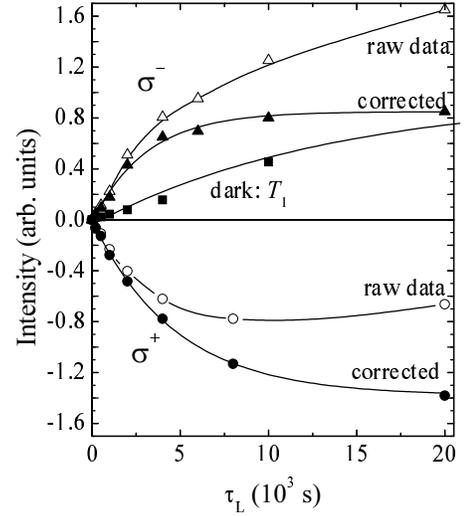}
\caption{\label{exposure}
Exposure time ($\tau_{\rm L}$) dependences of the integrated $^{31}$P signal intensity
with $\Phi$=430 mW/cm$^2$ at 4.2 K and 6.347 T.
The data for the dark case are also plotted.
The closed symbols represent the data corrected for the contribution from the dark part of the sample, 
and the solid lines are single exponential fits.}
\end{figure}

\begin{table}[t]
\caption{$E_p$ and helicity dependences of the buildup times of $^{31}$P 
under the light irradiation with $\Phi=$ 430 mW/cm$^2$ at 4.2 K and 6.347 T. 
The contribution from the dark part has been corrected.} 
\label{table:2}
\begin{tabular}{@{\hspace{\tabcolsep}\extracolsep{\fill}}ccc} \hline \hline
$E_p$ (eV) & Helicity & $T_b$ ($\times 10^3$s)\\ \hline
1.420	& $\sigma^{+}$ & 2.9 $\pm$ 0.2 \hspace{2mm} \\ \cline{2-3}
	& $\sigma^{-}$ & 2.6 $\pm$ 0.2 \hspace{2mm} \\ \hline
1.416	& $\sigma^{+}$ & 4.7 $\pm$ 0.2 \hspace{2mm} \\ \cline{2-3}
	& $\sigma^{-}$ & 4.6 $\pm$ 0.2 \hspace{2mm} \\ \hline \hline
\end{tabular}
\end{table}
Figure \ref{exposure} shows the exposure time ($\tau_{\rm L}$) dependence of the integrated $^{31}$P signal intensity
for $\sigma^{+}$ at 1.416 eV and for $\sigma^{-}$ at 1.420 eV,
where the maximal enhancement is observed for each helicity.
Since the light is illuminated only at a part of the sample, 
the observed signal is the sum of the contributions from the illuminated and the dark regions of the sample. 
The real $\tau_{\rm L}$ dependence in the illuminated region can be derived 
by subtracting that in the dark from the observed data,
which are shown by closed triangles and circles in Fig. \ref{exposure}.
The buildup times ($T_b$) obtained by single exponential fits to the corrected data
are shown for the two different $E_p$ and the helicities in Table \ref{table:2}.
One finds that the buildup time is independent of the helicity 
within the experimental error.

The maximal enhancement factor of the $^{31}$P signal can be estimated from Fig. \ref{exposure} 
by taking into account the buildup time as well as the volume of the illuminated region.
Here, we define the normalized signal enhancement $\epsilon$ as follows.
\begin{equation}
 \epsilon(E_p, \sigma^{\pm}, \tau_{\rm L}) 
\equiv \frac{I(E_p, \sigma^{\pm}, \tau_{\rm L})-I_{\rm BG}(\tau_{\rm LD}=\tau_{\rm L})}{I_{\rm BG}(\tau_{\rm LD}=\tau_{\rm L})},
\label{enhancement}
\end{equation}
where $I(E_p, \sigma^{\pm}, \tau_{\rm L})$ and $I_{\rm BG}(\tau_{\rm LD})$ are 
the integrated signal intensity measured 
under the light irradiation with $E_p, \sigma^{\pm}$ and $\tau_{\rm LD}$, 
and that of the background measured in the dark with $\tau_{\rm LD}=\tau_{\rm L}$, respectively.
The subtraction of $I_{\rm BG}(\tau_{\rm LD})$ in the numerator allows us to extract only the contribution 
from the illuminated region of the sample.
From Fig. \ref{exposure}, $\epsilon(1.416 eV, \sigma^+, \tau_{\rm L} \rightarrow \infty)=-1.4$,
while the area of the illuminated spot is 0.4 of the total area of the sample. 
Since the penetration depth of the light at 1.42 eV is 2 $\mu$m \cite{michal99},
the total volume where the optically pumped nuclei are involved is about 0.2 \% of the whole sample.
Hence, the enhancement factor for $E_p=$ 1.416 eV and $\sigma^{+}$ is  
$-1.4/(2 \times 10^{-3}) \sim - 7 \times 10^2$. 
Since $p$ of $^{31}$P at thermal equilibrium at 4.2 K is 6.3 $\times 10^{-4}$,
the average polarization of $^{31}$P nuclei in the illuminated region is estimated to reach about $-$40 \%.

\subsubsection{\label{photon-energy}Photon energy dependence of the $^{31}$P signal enhancement}
\begin{figure}[t]
\includegraphics[scale=0.5]{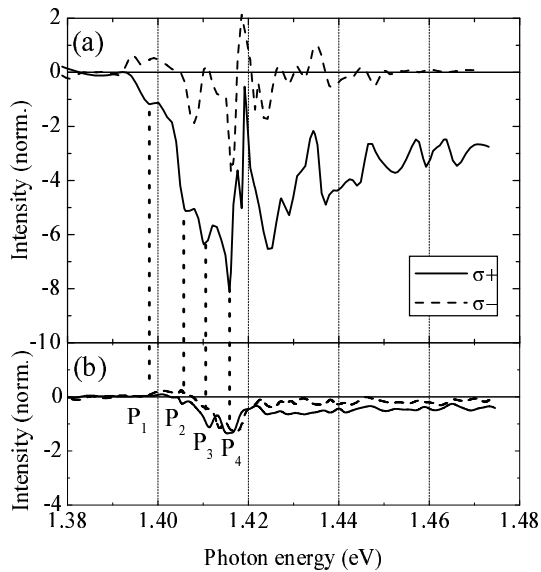}
\caption{\label{p-energy}
$E_p$ dependences of the enhancement $\epsilon(E_p, \sigma^{\pm}, \tau_{\rm L}=120$ s) 
for $^{31}$P with $\Phi$= 430 mW/cm$^2$ at 4.2 K. 
(a) $H_0$=6.347 T and (b)11.748 T.
P$_i$'s ($i=1 \sim 4$) indicate local minimum points (local maximal absolute values) 
of the enhancements for $\sigma^+$.
}
\end{figure}
Figure \ref{p-energy}(a) shows the $E_p$ dependences of the $^{31}$P signal enhancements 
$\epsilon(E_p, \sigma^{\pm}, \tau_{\rm L}=120$ s) $\equiv \epsilon(E_p, \sigma^{\pm})$ 
with $\Phi$= 430 mW/cm$^2$ at 4.2 K and 6.347 T.
Here, one can see oscillatory behaviors of $\epsilon(E_p, \sigma^{\pm})$ as a function of $E_p$. 
One may also notice that $\epsilon(E_p, \sigma^{\pm})$ is quite asymmetric in terms of helicity, 
i.e., $\epsilon(E_p, \sigma^{+}) \neq - \epsilon(E_p, \sigma^{-})$.
In particular, $\epsilon$(1.416 eV, $\sigma^-$) is negative
as is $\epsilon$(1.416 eV, $\sigma^+$).
These behaviors cannot be accounted for by the picture given in \S \ref{theory}.

We also performed the same experiment at 11.748 T, which is shown in Fig. \ref{p-energy} (b).
One can see that the absolute value of the enhancement is reduced by a factor of 4
from that in Fig. \ref{p-energy} (a).
The factor ``4'' is consistent with Eqs. (\ref{TII-2}) and (\ref{steady2})
in that $1/T_{II} \propto \omega_S^{-2} = (\gamma_e H_0)^{-2}$
and that $H_0$ is increased by 1.85 from 6.347 T to 11.748 T.
Note that the field dependence of the signal intensity irrelevant to the optical pumping effect 
is canceled out by the denominator $I_{\rm BG}$ in Eq. (\ref{enhancement}).
On the other hand, the oscillation period as a function of $E_p$ is the same for both the fields,
i.e., the local maxima and the minima appear at the same $E_p$'s.

\subsubsection{\label{op-115In}Photon energy dependence of the $^{115}$In signal enhancement}
\begin{figure}[t]
\includegraphics[scale=0.5]{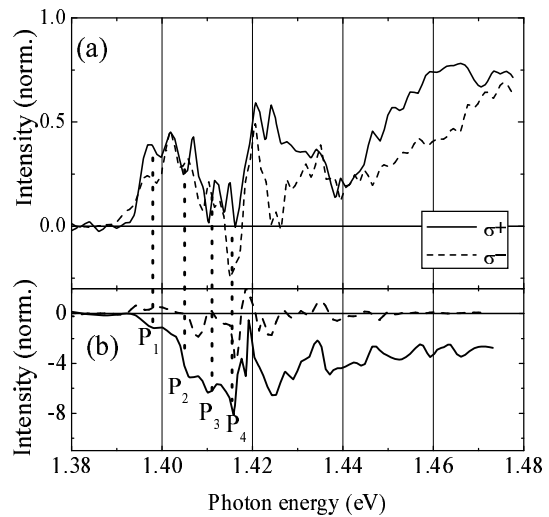}
\caption{\label{In-p-energy}
(a) $E_p$ dependences of the enhancement $\epsilon(E_p, \sigma^{\pm}, \tau_{\rm L}$=120 s) 
for $^{115}$In with $\Phi$= 430 mW/cm$^2$ at 4.2 K and 6.347 T.
(b) Corresponding $E_p$ dependence of $^{31}$P (Fig. \ref{p-energy} (a)).
The P$_i$'s are the same as those in Fig. \ref{p-energy} (a).}
\end{figure}
Fig. \ref{In-p-energy}(a) shows the $E_p$ dependences of the enhancement 
$\epsilon(E_p, \sigma^{\pm}, \tau_{\rm L}=120$ s) for $^{115}$In at 4.2 K and 6.347 T with $\Phi$=430 mW/cm$^2$.
For comparison, the data for $^{31}$P [Fig. \ref{p-energy}(a)] are shown in Fig. \ref{In-p-energy} (b). 
Here, one can find the following features for the $^{115}$In signal enhancements in this figure.
(1) The signal enhancements are positive for both the $\sigma^+$ and $\sigma^-$ except for $\sigma^-$ at around P$_4$.
(2) The difference between $\sigma^+$ and $\sigma^-$ is less significant than that for $^{31}$P.
(3) The signal enhancements also show oscillatory behaviors as seen in the case of $^{31}$P.
(4) The oscillation period is almost the same between $^{115}$In and $^{31}$P, but they are out-of-phase, 
i.e., the local minima for $^{115}$In indicated by P$_i$ coincide with the negative maxima for $^{31}$P.
Again, these features cannot be explained by the theory in \S \ref{theory}, 
indicating that the enhancement may be subject not only 
to the spin-selective interband transitions and hyperfine couplings, but also to some other mechanism.

A rather unexpected behavior has been also observed in the buildup times.
Table \ref{table:3} shows the buildup times for $^{31}$P and $^{115}$In at 4.2 K with $E_p$ = 1.420 eV
along with the $T_1$ values at 300 K.
One can see that the buildup times are of the same order between $^{115}$In and $^{31}$P.
This result is unexpected because the $T_1$ values are different 
by more than three orders of magnitude between them.
To see how it is unexpected, 
let us estimate the ratio of the buildup times between $^{115}$In and $^{31}$P (i.e., $^{31}T_{II}/^{115}T_{II}$)
expected from that of the $T_1$ values.
From Eq. (\ref{TII-2}),
\begin{equation}
^{31}T_{II}/^{115}T_{II}= (^{115}A/^{31}A)^2,
\end{equation}
while $(^{115}A/^{31}A)^2$ can be estimated from the $T_1$ values 
because $1/T_1 \propto I(I+1) A^2$. 
\begin{equation}
\left (\frac{^{115}A}{^{31}A} \right )^2=\frac{^{31}{\rm I}(^{31}{\rm I}+1)^{31}T_1}{^{115}{\rm I}(^{115}{\rm I}+1)^{115}T_1},
\end{equation}
where $^{31}$I=1/2 and $^{115}$I=9/2 are the nuclear spins for $^{31}$P and $^{115}$In, respectively.
From Table \ref{table:3}, 
\begin{equation}
^{31}T_1/^{115}T_1 \sim 1.5 \times 10^3,
\end{equation} 
so that $[^{115}A/^{31}A]^2$ is estimated to be $\sim 45$.
This is inconsistent with the fact that $T_{II}$ are of the same order for $^{31}$P and $^{115}$In.
We will address these unexpected results in the next section.
\begin{table}[t]
\caption{Buildup times for $^{31}$P and $^{115}$In with $E_p$ = 1.420 eV at 4.2 K and 6.347 T. 
The contribution from the dark part has been corrected.
Also shown are the spin-lattice relaxation times ($T_1$) of $^{31}$P and $^{115}$In at 300 K.
} 
\label{table:3}
\begin{tabular}{@{\hspace{\tabcolsep}\extracolsep{\fill}}cccc} \hline \hline
& & $^{115}$In & $^{31}$P \\ \hline
$T_b$ (s) & $\sigma^{+}$ & $1.6 \pm 0.4 \times 10^3$ \hspace{2mm} & $2.9 \pm 0.2 \times 10^3$ \\ \cline{2-4}
 	   & $\sigma^{-}$ & $1.4 \pm 0.5 \times 10^3$ \hspace{2mm} & $2.6 \pm 0.2 \times 10^3$ \\ \hline
$T_1$ (s) & - & 1.0 $\pm$ 0.1 $\times 10^{-1}$ \hspace{2mm} & 5.1 $\pm$ 0.2  $\times 10^2$\\ \hline \hline
\end{tabular}
\end{table}

\section{\label{discussion}Discussion}
The most peculiar aspect of the optical pumping effects in InP:Fe 
is the oscillatory behaviors of the signal enhancements as a function of $E_p$.
Michal et al. discussed its possible relation with the electron momentum relaxations $\tau_p$ 
caused by LO-phonons with discrete energy levels, 
but they found that the oscillation period did not match that estimated for the LO-phonons.\cite{michal99} 
Since this behavior is not observed in the undoped InP,\cite{farah98}
some unique properties of the InP:Fe sample may be responsible for it.
We speculate, from the data shown in the previous sections, 
that it stems from the structures of the photon absorption spectrum rather than relaxations.

Here, we divide this oscillation into two parts; i.e., the regions below and above P$_4$ in Fig. \ref{p-energy}.
The oscillation below P$_4$ may be caused by traps associated with phosphorus vacancies and Fe-ions.
The former exists even in the undoped InP, while the latter results from Fe-doping,
which compensates the electrons caused by the former.
In fact, the peaks in this region can be assigned from the photoluminescence (PL)
peaks reported in Ref. \onlinecite{kuriyama94}.
For example, $E_p$= 1.417 eV, labeled as P$_4$ can be assigned to the band edge transition,
while P$_1$ ($E_p$=1.396 eV) to excitons bound to deep level acceptors.
The two peaks between P$_1$ and P$_4$ (P$_2$, $E_p$=1.407 eV and P$_3$, 1.412 eV) 
have been also observed in the PL spectrum,
which may be related to excitons bound to neutral donors and/or acceptors.\cite{kuriyama94}

The oscillation above P$_4$, on the other hand, may be associated with 
the defect-acceptor or donor-acceptor pairs (DAP) in InP:Fe.\cite{kuriyama94} 
The process of a photo excitation in a DAP is described as,
\begin{equation}
D^+ + A^- + E_p \rightarrow D^0 + A^0,
\end{equation}
and the transition energy $E_p$ is given by,
\begin{equation}
E_p=E_g  -E_A- E_D + e^2/(\epsilon R_m),
\label{DAP-energy}
\end{equation}
where $E_g$, $E_A$, and $E_D$ are the energy for the gap energy and 
the binding energies in a neutral donor and an acceptor (D$^0$ and A$^0$), respectively.
The last term in Eq. (\ref{DAP-energy}) represents the coulomb binding energy ($E_B$) 
between the donor and the acceptor,
where $R_m$ is a distance from the donor of interest to the acceptor at the $m$-th nearest neighbor shell 
and $\epsilon$ is a dielectric constant, which is 9.6 in InP.\cite{sahu91}
Since the donor and the acceptor sites are localized at lattice points,
$R_m$ can take only discrete values determined by the crystal structure and the lattice constant $d$.
This results in discrete $E_B$ levels, and $E_p$ shows a series of discrete transition peaks.
The optical pumping effects with different $E_p$ occur in different DAPs,
and these discrete levels manifest themselves as the oscillatory behavior of the NMR signal enhancement
through $\tau_e$ in Eq. (\ref{TII}).
The energy difference in $E_B$ between adjacent shells 
($\Delta E_B \equiv E_B(R_m)-E_B(R_{m-1})$) falls within the range 
from 1 to 10 meV for $m=10 \sim 60$,\cite{thomas64}
which are comparable to the oscillation period of the signal enhancements
above P$_4$ in Figs. \ref{p-energy} and \ref{In-p-energy}.
We expect that the density of such accepter and donor pairs could be of the order of $10^{15}$ cm$^{-3}$,
because the undoped InP sample contains donor sites of the order of $10^{15}$ cm$^{-3}$, 
which are compensated by acceptors caused by the Fe-doping.

Let us turn to the issue of the $\sigma^{\pm}$ asymmetry 
manifested in Fig. \ref{p-energy}.
The asymmetry indicates that the polarization $p$ 
is not directly determined by $S_i$ in Eq. (\ref{S}).
On the other hand, the buildup time $T_b$ shown in Table \ref{table:2} 
is almost $\sigma^{\pm}$ independent in spite of the large $\sigma^{\pm}$ asymmetry of the enhancements.
These facts suggest that, besides the usual optical pumping process by photons,
there may be another process for buildup of the $^{31}$P polarization,
which is independent of the photon helicity.
The buildup process may be described phenomenologically (neglecting $T_1$ in the dark) as,
\begin{equation}
 p(\tau_{\rm L}) \propto p_1 (1-{\rm e}^{-\tau_{\rm L}/T_{II}})+p_2 (1-{\rm e}^{-\tau_{\rm L}/T_{I2}}),
\label{reorientation}
\end{equation}
where $T_{II}$ and $T_{I2}$ are the cross relaxation times of $^{31}$P 
by trapped electrons and by some unknown (second) process, respectively, 
and $p_1$ and $p_2$ are the corresponding polarizations at $\tau_{\rm L} \rightarrow \infty$.
Here, $p_1$ is negative (positive) for $\sigma^+$ ($\sigma^-$),
while $p_2$ is always negative regardless of $\sigma^{\pm}$.
Note that $p_1$, $p_2$, $T_{II}$ and $T_{I2}$ may be all $E_p$ dependent. 
If $T_{I2} \gg T_{II}$, $T_b$ is predominated by the second process.

A plausible mechanism for this second process is 
that caused by the dipolar ordered $^{113/115}$In nuclei,
as manifested by the single resonance polarization transfer effect.\cite{michal98}
This effect is caused by the Hartmann-Hahn cross-polarization mechanism 
between $^{31}$P nuclei in the rotating frame 
and dipolar ordered $^{113/115}$In nuclei in the {\it laboratory} frame.
In the dipolar ordered state, 
$^{113/115}$In nuclei form domains 
in which the nuclei are correlated by nuclear dipolar couplings.
\cite{anderson62}
The energy required to flip a \textit{spin} of the domain 
is not only a Zeeman energy of one nucleus $\hbar \omega_S$, 
but also of the order of nuclear dipolar-dipolar interactions 
because of the strong nuclear spin correlations 
(in other words, multiple quantum coherences) in each domain.
We may expect that a similar mechanism 
renders \textit{reorientation} of $^{31}$P by the dipolar ordered indium nuclei.

Here, one may pose a question: i.e., 
provided that $^{113/115}$In are responsible for the second process ($T_{I2}$), 
how can we expect the phenomenon similar to the single resonance nuclear polarization transfer,
although neither $^{31}$P nor $^{113}$In are in the resonance condition during the light exposure?
We consider that the reorientation of $^{31}$P 
is caused by the dynamic nuclear self-polarization, 
where $^{31}$P nuclei interact indirectly with $^{113/115}$In nuclei via electrons.
\cite{dyakonov72,meier84,farah98}
The mechanism is as follows.
The polarization of nuclei created by the optical pumping produces nuclear magnetic fields for electron spins,
which causes net electron spin $S_z$ in Eq. (\ref{S}) to change.
Simultaneously, the change in $S_z$ gives rise to the change in the nuclear polarization
as given by Eq. (\ref{steady1}), 
and as a result, the interactions among $^{113}$In, $^{115}$In and $^{31}$P are set up.

Note that this process is different from the usual indirect nuclear spin-spin coupling,
\textit{e.g.} the Suhl-Nakamura interaction in magnets, 
where electron spins appear in the process only implicitly as a virtual process, 
so that the energy conservation between the nuclear spins is strictly observed.
In the present process, on the other hand, changes in the electron spins 
are not virtual but \textit{actual}.
The influence of the electron spins on the nuclei manifests itself as a static internal field,
and the difference in the Zeeman energies between $^{31}$P and $^{113/115}$In 
could affect only a dynamical aspect of the process through $T_{II}$ given by Eq. (\ref{TII}).
In the present case, however, $T_{I2} \gg T_{II}$, so that 
$T_{II}$ of $^{31}$P may not be a determinant of $T_b$.
Since the process proceeds in a cooperative way among $^{113}$In, $^{115}$In and $^{31}$P nuclei
via the net electron spins $S_z$,
the same order of $T_b$ is anticipated for $^{115}$In and $^{31}$P, 
which is what we have actually seen in Table \ref{table:3}.

We believe that these mechanisms are rather plausible.
Nevertheless, further investigations are needed to elucidate these peculiar behaviors.
\vspace{.5cm}

\section{\label{sec:conclusion}Conclusion}
We have investigated the optical pumping NMR in InP:Fe,
which exhibits the most intense optical pumping effect among other dopants.
The effect has been optimized 
and the $^{31}$P polarization up to about 40 \% has been achieved
with $\sigma^{+}$ and $E_p$ = 1.416 eV.

The quite effective optical pumping effect in InP:Fe is accompanied by some peculiar phenomena 
such as the oscillatory behavior of the NMR signal enhancement as a function of $E_p$
and the asymmetry of the enhancement against $\sigma^{\pm}$.
We have revealed that the buildup time for the $^{31}$P enhancement is almost independent of the helicity,
and has the same order as that of $^{115}$In.
These results indicate that the buildup process of $^{31}$P polarization 
is not determined by the usual optical pumping process by photon absorption. 
We have discussed the possible reorientation of the $^{31}$P nuclei due to the indirect couplings 
with the dipolar ordered $^{113/115}$In polarizations via trapped electrons.
On the other hand, the magnetic-field and nucleus ($^{31}$P and $^{115}$In) 
independence of the oscillation period suggests that 
the oscillation is caused by traps at phosphorus vacancies and/or donor-acceptor pairs (DAP), 
which provide a discrete photo-excitation spectrum.

The present work has shown that the optical pumping NMR is an effective tool 
for the investigations of impurity levels in semiconductors.
 
\section*{Acknowledgments}
The authors would like to acknowledge helpful advice by M. Oshikiri.
One of the authors (A. G.) is also indebted to R. Tycko for helpful discussion.
This work has been partially supported by Industrial Technology Research Grant Program in '03 
from New Energy and Industrial Technology Development Organization (NEDO) of Japan.

%\newpage %Just because of unusual number of tables stacked at end
\bibliography{inp-opnmr}% Produces the bibliography via BibTeX.

\end{document}